\documentclass[useAMS,usenatbib]{mn2e}
\usepackage{graphicx}
\usepackage{lscape}
\usepackage{listings}
\usepackage{url}
\usepackage{array}
\usepackage{float}

\usepackage{color}
\usepackage{listings}

\newfloat{Query}{H}{myc}

\title[A T6.5 subdwarf]{The discovery of a T6.5 subdwarf}
\author[Ben Burningham et al]{Ben Burningham$^{1}$\thanks{E-mail:
    B.Burningham@herts.ac.uk},  L. Smith$^{1}$, C. V. Cardoso$^{2,1}$, P.W. Lucas$^{1}$, A. J. Burgasser$^{3}$, 
\newauthor
H. R. A. Jones$^{1}$, R. L. Smart$^{2}$ \\
$^{1}$ Centre for Astrophysics Research, Science and Technology Research Institute, University of Hertfordshire, Hatfield AL10 9AB, UK\\
$^{2}$ Istituto Nazionale di Astrofisica, Osservatorio Astrofisico di Torino, Strada Osservatorio 20, 10025 Pino Torinese, Italy \\
$^{3}$ Center for Astrophysics and Space Science, University of California San Diego, La Jolla, CA 92093, USA \\
}

\begin{document}
%
%
%
%


\def\aj{\rm{AJ}}                   
\def\araa{\rm{ARA\&A}}             
\def\apj{\rm{ApJ}}                 
\def\apjl{\rm{ApJ}}                
\def\apjs{\rm{ApJS}}               
\def\ao{\rm{Appl.~Opt.}}           
\def\apss{\rm{Ap\&SS}}             
\def\aap{\rm{A\&A}}                
\def\aapr{\rm{A\&A~Rev.}}          
\def\aaps{\rm{A\&AS}}              
\def\azh{\rm{AZh}}                 
\def\baas{\rm{BAAS}}               
\def\jrasc{\rm{JRASC}}             
\def\memras{\rm{MmRAS}}            
\def\mnras{\rm{MNRAS}}             
\def\pra{\rm{Phys.~Rev.~A}}        
\def\prb{\rm{Phys.~Rev.~B}}        
\def\prc{\rm{Phys.~Rev.~C}}        
\def\prd{\rm{Phys.~Rev.~D}}        
\def\pre{\rm{Phys.~Rev.~E}}        
\def\prl{\rm{Phys.~Rev.~Lett.}}    
\def\pasp{\rm{PASP}}               
\def\pasj{\rm{PASJ}}               
\def\qjras{\rm{QJRAS}}             
\def\skytel{\rm{S\&T}}             
\def\solphys{\rm{Sol.~Phys.}}      
\def\sovast{\rm{Soviet~Ast.}}      
\def\ssr{\rm{Space~Sci.~Rev.}}     
\def\zap{\rm{ZAp}}                 
\def\nat{\rm{Nature}}              
\def\iaucirc{\rm{IAU~Circ.}}       
\def\aplett{\rm{Astrophys.~Lett.}} 
\def\apspr{\rm{Astrophys.~Space~Phys.~Res.}}
\def\bain{\rm{Bull.~Astron.~Inst.~Netherlands}} 
\def\fcp{\rm{Fund.~Cosmic~Phys.}}  
\def\gca{\rm{Geochim.~Cosmochim.~Acta}}   
\def\grl{\rm{Geophys.~Res.~Lett.}} 
\def\jcp{\rm{J.~Chem.~Phys.}}      
\def\jgr{\rm{J.~Geophys.~Res.}}    
\def\jqsrt{\rm{J.~Quant.~Spec.~Radiat.~Transf.}}
\def\memsai{\rm{Mem.~Soc.~Astron.~Italiana}}
\def\nphysa{\rm{Nucl.~Phys.~A}}   
\def\physrep{\rm{Phys.~Rep.}}   
\def\physscr{\rm{Phys.~Scr}}   
\def\planss{\rm{Planet.~Space~Sci.}}   
\def\procspie{\rm{Proc.~SPIE}}   

\let\astap=\aap
\let\apjlett=\apjl
\let\apjsupp=\apjs
\let\applopt=\ao

\maketitle

\begin{abstract}
We report the discovery of ULAS~J131610.28+075553.0, a sdT6.5 dwarf in the UKIDSS Large Area Survey 2 epoch proper motion catalogue. This object displays significant spectral peculiarity, with the largest yet seen deviations from T6 and T7 templates in the $Y$ and $K$ bands for this subtype. Its large, $\sim 1$~arcsec/yr,  proper motion suggests a large tangential velocity of $V_{tan} \approx 240 - 340$kms$^{-1}$, if we assume its $M_J$ lies within the typical range for T6.5 dwarfs. This makes it a candidate for membership of the Galactic halo population. However, other metal poor T dwarfs exhibit significant under luminosity both in specific bands and bolometrically. As a result, it is likely that its velocity is somewhat smaller, and we conclude it is a likely thick disc or halo member. This object represents the only T dwarf earlier than T8 to be classified as a subdwarf, and is a significant addition to the currently small number of known unambiguously substellar subdwarfs. 
\end{abstract}


\section{Introduction}
\label{sec:intro}
The current generation of wide field surveys are bringing about a step change in our understanding of the coolest and lowest mass components of the Solar neighbourhood.
The total number of cool T dwarfs, substellar objects with 1400~K$> T_{\rm eff} > 500$K, has been taken into the hundreds by infrared surveys such as the Sloan Digital Sky Survey \citep[SDSS; ][]{sdss}, the 2 Micron All Sky Survey \citep[2MASS; ][]{2mass}, the UKIRT Infrared Deep Sky
Survey \citep[UKIDSS; ][]{ukidss}, the Canada-France Brown Dwarf Survey \citep[CFBDS; ][]{cfbds} and most recently the Wide field Infrared Survey Explorer \citep[WISE; ][]{wise}. The last of these, which is
an all-sky mid-infrared survey, has extended the substellar census to well below Teff = 500 K, and the adoption of a new spectral class ÒYÓ has been suggested to classify these
new extremely cool objects \citep{cushing2011,kirkpatrick2012}.This rapid expansion of local brown dwarf census is revealing a growing number of objects that lie at the extremes of the parameter space covered by substellar and planetary atmospheric model grids. 

For example, in addition to the extremely cool brown dwarfs discovered by WISE, several young late-type T~dwarfs are now known with surface gravities and temperatures that overlap with directly imaged exoplanets \citep[$\log g = 3.5 - 4.0$; $T_{\rm eff} \sim 700$~K; e.g.  ][]{goldman2010,burgasser2010,ben2011b,delorme2012,kuzuhara2013,thalmann2009,janson2011}. 
At the other extreme, a number of extremely high surface gravity T~dwarfs spanning the same temperature range have also been identified \citep[$\log g \sim 5.0 - 5.3$;  e.g. ][]{ben10a,pinfield2012}. The range of surface gravity probed in the sub-1000K regime thus now spans nearly 2 dex. 

By contrast, the range of metallicities that has been probed remains stubbornly narrow, reflecting the local Galactic disc population. So far, only one T~dwarf has yet been confirmed to have a metallicity beyond ${\rm [M/H]} \sim \pm 0.4$ \citep{mace2013b}, and only a handful of warmer L~dwarfs are known with halo kinematics and correspondingly low-metallicities \citep[e.g. ][]{burgasser03,lodieu10,burgasser2009}. 
This is not for lack of hunting.  The first T~dwarfs to be identified with subsolar metallicities are only moderately metal-poor \citep[$-0.4 < {\rm [M/H]} < 0.0$; ][]{BBK06}, and have thin-disc kinematics.
A systematic search of the UKIDSS Data Release 5 identified just 2 candidate halo T dwarfs from a sample of approximately 100 objects. These candidates were modest ($2\sigma$) kinematic outliers from the background disc population \citep{murray2011}, and one (ULAS~J131943.77+120900.2) has since been ruled out upon re-measurement of its proper motion \citep{smith2014}. More recently, \citet{pinfield2013} has identified two T8--T9 dwarfs with likely thick-disc/halo kinematics, one of which displays similar spectral peculiarity to the T8 subdwarf identified by \citet{mace2013b}.

Despite this relatively narrow range currently probed, even small shifts in metallicity have been shown to have significant impact on the observed properties of late-type T~dwarfs. For example, $0.3$~dex shifts in metallicity have as much impact on the $H - {\rm [4.5]}$ colours of T8 dwarfs as a 100~K shift in $T_{\rm eff}$ \citep{ben2013}. Identifying cool T dwarf members of the Galactic halo provides an opportunity to expand the currently observed parameter space, and provide new tests of atmospheric models. Furthermore,  identification of  a population of T~subdwarfs offers unambiguous confirmation that substellar objects can form in very low-metallicity environments, and will provide the starting point for exploring the substellar IMF of the halo and thick disc.

In this Paper we present the discovery of ULAS~J131610.28+075553.0 (ULAS~J1316+0755), a T6.5 subdwarf.  Sections~\ref{sec:ident}-\ref{sec:spec} deal with its initial identification and follow-up observations and Section~\ref{sec:kin} lays out a detailed discussion of its likely kinematic membership and the justification for the subdwarf classification.

\section{Initial identification}
\label{sec:ident}

Our strategy for identifying T~dwarfs in the UKIDSS Large Area Survey (LAS) has been described in detail in \citet{ben10b} and \citet{ben2013}.  As part of these searches we identified ULAS~J1316+0755 as a relatively blue ($Y- J = 0.75$) late-T dwarf candidate. Only detected in $YJ$, the candidate was undetected in UKIDSS images $HK$, and was undetected in SDSS DR8. 

\begin{figure}
\includegraphics[width=8cm,angle=0]{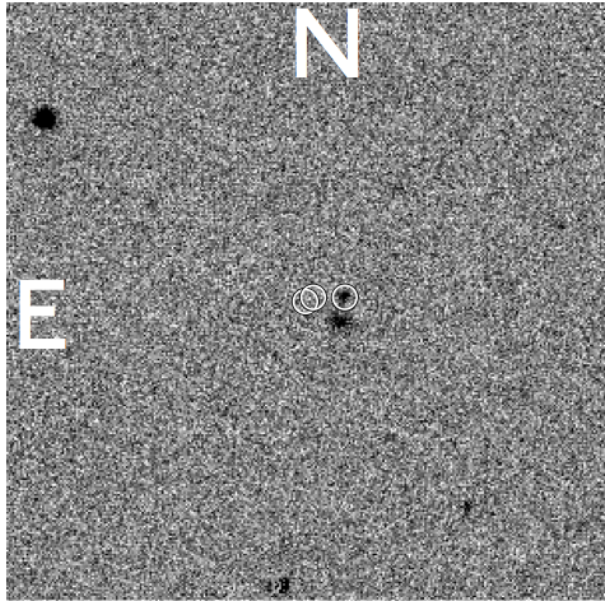}
\caption{The $J$ band image of ULAS~J1316+0755 from the UKIDSS LAS taken in 2010. Positions for the epochs 2006.4088, 2007.1516 and 2010.2366 show the target's motion from east to west. Each side is 1 arcminute long. }
\label{fig:finder}
\end{figure}

All candidate and confirmed T dwarfs were crossmatched against the new LAS 2 epoch proper motion catalogue of \citet{smith2014}, and proper motions of previously confirmed T~dwarfs have now been published in \citet{ben2013}.  As a result of this crossmatch, ULAS~J1316+0755 was highlighted as the highest proper motion target in our sample.  ULAS~J1316+0755 was observed three times in the $J$ band in 2006, 2007 and 2010. A finder chart from the 2010 image is shown in Figure~\ref{fig:finder}, and 2006, 2007 and 2010 positions for ULAS~J1316+0755 are overlaid.  The photometric and astrometric properties of ULAS~J1316+0755 are summarised in Table~\ref{tab:properties}.

\begin{figure}
\includegraphics[width=8cm,angle=0]{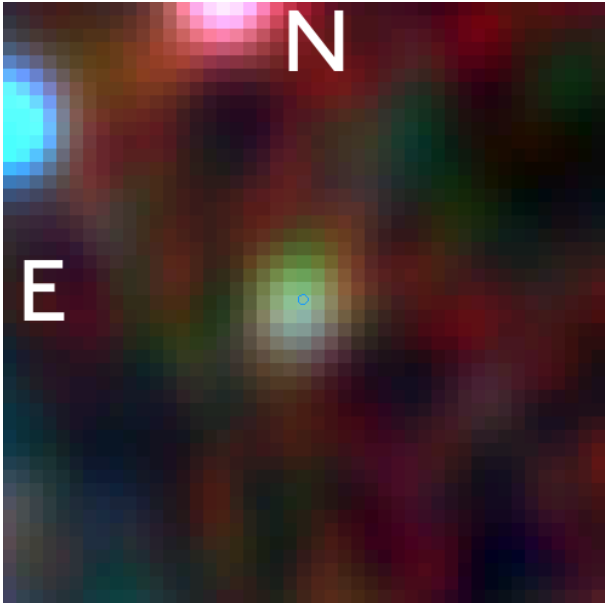}
\caption{A false colour WISE 1 arc minute square image from 2010.5 centred on the 2010.2 coordinates of ULAS~J1316+0755.  W1, W2 and W3 are indicated by blue, green and red respectively. }
\label{fig:wise}
\end{figure}

Figure~\ref{fig:wise} shows a false colour image from the WISE All Sky Release \citep{wise}. Unfortunately, ULAS~J1316+0755 is blended with the background galaxy to its south in the WISE images, making interpretation of its photometry problematic. The green (W2) extension to the northern edge of the WISE source suggests that ULAS~J1316+0755 has been detected and contributes significant flux in this filter.
The recent {\it AllWISE} data release identifies a source at this location with  $W2 = 15.50 \pm 0.1$, and proper motion $mu_{\alpha \cos \delta}  = -684 \pm 458$mas/yr, $\mu_{\delta} = -318 \pm 503$mas/yr.  These astrometric values are consistent with the UKIDSS ones albeit with error bars more than an order of magnitude higher, and support the assertion that the moving source is not entirely swamped by light from the background galaxy.
  The blended source has $W1 - W2 = 1.1 \pm 0.1$, $J-W2 = 4.1 \pm 0.1$ (using the UKIDSS $J$ band photometry for the point source),  and $W2-W3 = 3.1 \pm 0.3$. The first two of these are consistent with expectations for a T dwarf, but the $W2 -W3$ colour is too red for inclusion in the T~dwarf searches carried out by the WISE brown dwarf team \citep[e.g.][]{kirkpatrick2011,mace2013}, and reflects the impact of the extended extragalactic source.
It is thus apparent that the flux contributed by ULAS J1316+0755 is too blended with the galaxy to allow the extraction of useful photometry.  

\section{Methane photometry}
\label{sec:ch4}

Narrow band methane photometry was obtained on 2013 January 26 (UT) with NICS, the Near Infrared Camera Spectrometer \citep{baffa2001} mounted at the Cassegrain focus of the 3.5m Telescopio Nazionale Galileo (TNG) on Roque de Los Muchachos Observatory (ORM, La Palma, Spain). NICS is equiped with a 1024x1024 Rockwell HAWAII-1 HgCdTe that comprises a $4.2 \times 4.2 \arcmin$ field of view.
The observations were taken under program AOT26 TAC68.

The Speedy Near-infrared data Automatic Pipeline (SNAP) provided by TNG (version 1.3) was used to perform flat-fielding, compute the offsets between the dithered images, correct for field distortion and to create the final mosaic images with double-pass sky subtraction.

The target was observed for 60 minutes in $\rm{CH{_4}l}$, using 30 s exposures with 4 coadds and a 30 pointing dithering, and for 39 minutes in $\rm{CH{_4}s}$, using 26 s exposures with 3 coadds and a 30 pointing dithering. The seeing during the observations was approximately $0.9 \arcsec$.
The photometric extraction was performed with IMCORE part of CASUTOOLS using a fixed circular aperture of $2 \arcsec$ radius.

Methane differential photometric colour can be used as a proxy for spectral typing for T brown dwarfs. To calibrate the methane photometric system we have used the method defined by \cite{tinney2005} and revised for T brown dwarfs later than T3 by Cardoso et al. (submitted).
For ULASJ1316+0755 we have obtained a methane colour of $\rm{CH{_4}s} - \rm{CH{_4}l} = -0.45 \pm 0.08$, which would be the expected value for a T brown dwarf with a spectral type of a $\rm{T}4.8^{+0.6}_{-0.8}$.

\begin{table}
\begin{tabular}{ c c }
\hline
Name &  ULAS~J131610.28+075553.0 \\
$\alpha_{J2000}$ (epoch 2006.41) & 13:16:10.28\\
$\delta_{J2000}$ (epoch 2006.41) &  +07:55:53.0\\
$\mu_{\alpha \cos \delta}$ & $-1012.2 \pm 15.24$~mas/yr\\
$\mu_{\delta}$ & $+102.8 \pm 13.85$~mas/yr\\
$Y_{MKO}$(2006.41) & $20.00 \pm 0.14$ \\
$Y_{MKO}$(2007.15) & $20.04 \pm 0.15$ \\
$Y_{MKO}$(2010.24) &  $19.73 \pm 0.14$\\
$J_{MKO}$(2006.41) &  $19.29 \pm 0.12$ \\
$J_{MKO}$(2007.15) &  $19.21 \pm 0.11$\\
$J_{MKO}$(2010.24) &  $18.98 \pm 0.10$\\
CH$_{4} s - l$ & $-0.45  \pm 0.13$ \\
Spectral type & sdT6.5$ \pm 0.5$ \\ 
\hline
\end{tabular}
\caption{The measured properties of ULAS~J1316+0755. See Sections~\ref{sec:spec} and~\ref{sec:kin} for discussion of this object's spectral type.
\label{tab:properties}
}
\end{table}

\section{Spectroscopic confirmation}
\label{sec:spec}

We obtained follow-up spectroscopy of ULAS~J1316+0755 using the Gemini Near Infrared spectrograph \citep[GNIRS;][]{gnirs} on the Gemini North Telescope\footnote{under program GN-2013A-DD-2} on the night of  2013 April 13 (UT).
The observations were made up of a set of 300~second sub-exposures in an ABBA jitter pattern to facilitate effective background subtraction, with a slit width of 1 arcsec. The length of the A-B jitter was 10 arcsecs and the pattern was repeated 3 times to give a total integration time of one hour. 

The observations were reduced using standard IRAF
Gemini packages {\citep{cooke2005}. 
Comparison argon arc frames were
used to obtain dispersion solutions, which were then applied to the
pixel coordinates in the dispersion direction on the images.
The resulting wavelength-calibrated subtracted pairs had a low-level
of residual sky emission removed by fitting and subtracting this
emission with a set of polynomial functions fit to each pixel row
perpendicular to the dispersion direction, and considering pixel data
on either side of the target spectrum only. 
The spectra were then extracted using a linear aperture, and cosmic
rays and bad pixels removed using a sigma-clipping algorithm.
Telluric correction was achieved by dividing each extracted target
spectrum by that of  the F7V star LTT 14284, which was observed just 
after the target and at a similar airmass.
Prior to division, hydrogen lines were removed from the standard star
spectrum by interpolating the stellar continuum.
Relative flux calibration was then achieved by multiplying through by a
blackbody spectrum with a $T_{\rm eff} = 6200$K.

We have also obtained additional spectroscopy for the T6p low-metallicity benchmark HIP 73786B (Murray et al. 2011), to provide a metallicity-calibrated comparison of similar spectral type.  HIP 73786B was observed with the SpeX spectrograph \citep{spex} on the 3m NASA Infrared Telescope Facility
on 2011 April 19 (UT) in clear conditions.  The source was observed with SpeX's prism mode and 0$\farcs$5 slit, which provides 0.8--2.5~$\micron$ continuous spectroscopy with an average resolution $\lambda/\Delta\lambda \approx$ 120.  Eight exposures of 150~s each were obtained at an airmass of 1.05, dithering along the 15$\arcsec$ slit which was aligned with the parallactic angle.  We also observed the A0~V star HD~136831 ($V$ = 6.28) at a similar airmass for flux calibration, and obtained Ar arc and incandescent flat field lamps for wavelength and pixel response calibration. Data were reduced using SpeXtool \citep{vacca2003,cushing2004} using standard procedures.

In the top panel of Figure~\ref{fig:sptype} we have compared our GNIRS YJHK spectrum of ULAS~J1316+0755 to the T6 (SDSSp~J162414.37+002915.6) and T7 (2MASSI~J0727182+171001) spectral type standard templates \citep{burgasser06}. ULAS~J1316+0755 matches the T7 template best in the $J$~band, but the T6 in $H$ band. It matches neither of them in the $Y$ and $K$ bands, where it displays enhanced $Y$ band flux and depressed $K$ flux, characteristic of a high-gravity and low-metallicity atmosphere.  This suggests a classification of T6.5p may be appropriate, but we reserve our final classification for the end of Section~\ref{sec:kin}. This is a somewhat later classification than was suggested by the narrow-band methane photometry. The discrepancy is approximately one subtype beyond the estimated error range, which is fairly significant. We do not know the origin of this discrepancy, but repeat observations of this target would be useful to rule out periodic variability.

There is currently a dearth of atmospheric model grids at extremely low-metallicity, and the moderately low-metallicity models give poor fits to very low-metallicity object spectra \citep[e.g. ][]{mace2013}. However, the models that exist likely correctly predict qualitative trends in spectral morphology, such as $Y$ band enhancement and $K$ band suppression, even if quantitative predictions are not yet reliable. In trying to infer the properties of ULAS~J1316+0755 we thus rely on comparisons with benchmark low-metallicity objects, and well characterised non-benchmark objects, in combination with the qualitative expectations derived from available model grids  \citep[e.g. ][]{burrows2006, sm08, btsettlCS16}.

In the lower panel, we have compared ULAS~J1316+0755 to the spectrally peculiar T~dwarf SDSS~J141624.08+134826.7B (T7.5p) \citep[SDSS~J1416+1348B; ][]{ben10a}, which is thought to be low-metallicity and high gravity, and the new $YJHK$ Spex Prism spectrum of ${\rm [M/H]} \approx -0.3$ benchmark HIP~73786B \citep{murray2011}. All three objects show depressed $K$ band  and enhanced $Y$ band flux peaks. However, ULAS~J1316+0755 is noticeably more extreme than both the comparison objects in the $Y$~band, and much more depressed than HIP~73786B in the $K$~band.  Suppressed $K$ band emission is thought to be indicative of low-metallicity and/or high-gravity \citep{burgasser02, knapp04, liu07}, caused by pressure-enhanced collision induced absorption by hydrogen \citep[CIA H$_2$; ][]{saumon94}.
Meanwhile, the shape and height of the $Y$ band flux peak is thought to be less affected by gravity, and more indicative of metallicity, with a brighter and broader flux peak corresponding to lower metallicity, due to reduced opacity in the wings of the  0.77$\mu$m {\sc K I} line \citep[e.g. ][]{BBK06,burrows2006}. This suggests that ULAS~J1316+0755 is the lowest metallicity T6--T7 dwarf yet identified.

\begin{figure*}
\includegraphics[width=10cm,angle=90]{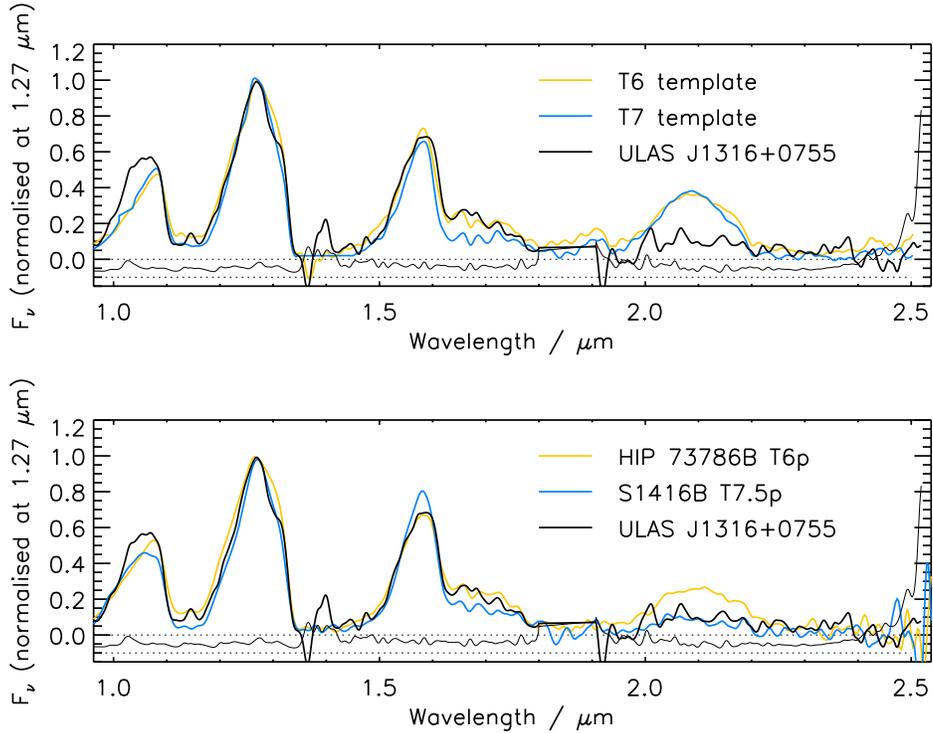}
\caption{{\bf Top panel:} The GNIRS $YJHK$ spectrum of ULAS~J1316+0755 along with the T6 (SDSSp~J162414.37+002915.6) and T7 (2MASSI~J0727182+171001) spectral type templates \citep{burgasser06}. {\bf Bottom panel: } The GNIRS $YJHK$ spectrum of ULAS~J1316+0755 compared to the previously identified metal poor T dwarfs, SDSS~J141624.08+134826.7B (T7.5p) and the metal-poor T6p benchmark HIP73786B \citep{murray2011}. For SDSS~J1416+1348B we have plotted the $YJHK$ Spex spectrum presented by \citet{burgasser2010}.}
\label{fig:sptype}
\end{figure*}

\section{The kinematic membership of ULAS~J1316+0755}
\label{sec:kin}
 
Its high proper motion, faint apparent magnitude and spectral peculiarities suggestive of low-metallicity make ULAS~J1316+0755 a strong candidate for membership of the Galactic halo. However, assessing its kinematic properties relies on estimating its distance effectively. In the absence of a parallax, this process is necessarily somewhat speculative for such an unusual object since the impact of metallicity on luminosity and specific magnitudes at a given spectral type is not yet well understood. 

Recently, \citet{mace2013b} have identified a T8 companion to the sdM1.5 star Wolf 1130. Wolf 1130C is considered the prototypical T8 subdwarf by association with the M subdwarf primary star.This system has thick-disc kinematics, and low-metallicity. Measuring metallicities for M~dwarfs is difficult due to the complex molecular opacity sources in their atmospheres and only two spectroscopic estimates for Wolf 1130 have been made. \citet{babs2012} report  ${\rm [M/H] = }-0.45 \pm 0.12$dex, and ${\rm [Fe/H] =}-0.64 \pm 0.17$dex for Wolf~1130 based on a calibration of $K$~band Na~{\sc I} K~{\sc I} equivalent widths and their H$_{2}$O--K2 index against M~dwarf wide-companions to FGK stars. \citet{woolf2006} used high resolution optical spectroscopy to measure the Iron abundance for Wolf 1130 directly, finding a similar value of  ${\rm [Fe/H] =}-0.62 \pm 0.10$dex, and estimated ${\rm [M/H] = }-0.52$~dex (no error quoted).

Wolf 1130C displays considerable spectral peculiarity and is also notably faint in several bands. In particular, it has $M_{J}(MKO) = 18.64$, 2.25 magnitudes fainter than the mean for T8 dwarfs reported by \citet{dupuy2013}, $M_{J}(MKO) =  16.43$. Similarly, the metal poor T dwarfs BD+01~2920B \citep[T8p, ${\rm [M/H] = -0.36}$; ][]{pinfield2012}, and SDSS~J1416+1348B (T7.5p) are both approximately 1 magnitude fainter than the means at their respective types \citep{pinfield2012,dupuy2012}.

At the T6 subtype, the impact of metallicity on $M_J$ is not clear, and the range of metallicities probed by benchmarks is smaller. Both the T6p benchmark, HIP~73786B (${\rm [M/H}] \approx -0.3$) and the suspected metal poor T6p dwarf 2MASS~J09373487+2931409 \citep[2MASS~J0937+2931; ][]{burgasser02,burgasser06} display $M_J$ very close to the mean for T6 dwarfs \citep{dupuy2012}, although their bolometric luminosities are both marginally lower than typical \citep{golim04,ben2013}.  This lack of reduction in $J$ band luminosity, compared to that seen in Wolf~1130C, may be the result of differences in the degree to which CIA H$_2$ absorption affects this region of the spectrum at different temperatures and/or some temperature dependence on how metallicity affects the $T_{\rm eff}$--spectral type relationship.  However, since comparison with (the benchmark) HIP~73786B suggests that ULAS~J1316+0755 has ${\rm [M/H]} < -0.3$, and we cannot rule out a metallicity even lower than Wolf~1130C, we must also consider the possibility that it may be fainter at $J$ than is typical for thin disc T6.5 dwarfs in our analysis.

If we allow for a $\pm 0.5$ subtype uncertainty, and the scatter about the mean $M_J$ for T6 and T7 dwarfs reported by Dupuy et al (2012), we estimate a spectrophotometric distance range of 50 -- 70~pc. This suggests tangential velocity of $V_{tan} \approx 240 - 340$kms$^{-1}$. Such a high tangential velocity would imply that this source is likely a member of the Galactic halo \citep[e.g. ][]{bensby2003}.  On the other hand, if we apply a +2.25 magnitude offset to the expected $M_{J}$ for ULAS~J1316+0755, corresponding to that seen in the case of Wolf~1130C, our spectrophotometric distance reduces to 19 -- 26~pc, with a corresponding $V_{tan} \approx 90 - 125$kms$^{-1}$. We have translated these two extremes of the likely tangential velocity estimate to $UV$ velocities, assuming a range of radial velocities, $V_{rad}$, and have plotted them in Figure~\ref{fig:uvplot}, along with a selection of Galactic populations from \citet{soubiran2008}. Figure~\ref{fig:uvplot} highlights the strong dependence of assigned kinematic family on the assumed distance for ULAS~J1316+0755.

\begin{figure}
\includegraphics[width=8cm,angle=90]{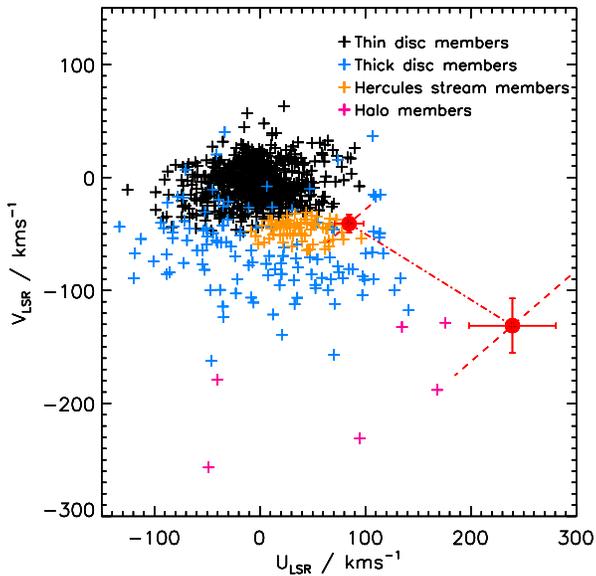}
\caption{Estimated UV velocities for ULAS~J1316+0755 based on our two extreme distance estimates (60~pc and 22.5~pc) are shown with large red symbols and associated error bars. The impact of  plausible radial velocities in the range  $V_{rad} = \pm V_{tan} / \sqrt{2}$~kms${-1}$ are indicated with red dashed lines. Velocities for a representative sample of Galactic stars from \citet{soubiran2008} are also plotted.}
\label{fig:uvplot}
\end{figure}

We have calculated ULAS~J1316+0755's probability of membership of the thin disc ($P_D$), thick disc ($P_{TD}$) and halo ($P_H$) by using the expressions for relative membership probabilities given by equations 1--3 of \citet{bensby2003}, normalised to give absolute probabilities under the assumption that $P_{D} + P_{TD} + P_{H} = 1$.  We have used the Solar neighbourhood fractions for thin disc, thick disc and halo stars given by \citet{bensby2003} and assumed three cases of $V_{rad} = 0, \pm (V_{tan}/\sqrt{2})$~kms$^{-1}$.
In Table~\ref{tab:probs} we list the probabilities of group membership for ULAS~J1316+0755 under three different assumptions of distance. In Figure~\ref{fig:memprob} we have represented these probabilities graphically.

\begin{table}
\begin{tabular}{c c c c c c}
\hline
Distance & $V_{tan}$ & $V_{rad}$  & $P_{D}$ & $P_{TD}$ & $P_H$ \\ 
pc & kms$^{-1}$ & kms$^{-1}$ & & & \\
\hline
	  &  & +205 &  0.0000 & 0.0000 & 1.00 \\
60 &    290 & 0    & 0.0000& 0.0011 & 0.9989 \\   
	&  & -205 & 0.0000 & 0.0000 & 1.0000\\
	  &&&&&\\
	 & & +130  & 0.0000 & 0.0107 & 0.9893 \\
38   & 180 & 0 &0.0004 & 0.9442 & 0.0554 \\
	 & & -130 & 0.0000 & 0.9175 & 0.0825 \\
	  &&&&&\\
	& & +76 & 0.0000 & 0.9843 & 0.0157 \\ 
22.5  & 110 & 0 & 0.7750 & 0.2248 & 0.0003 \\
	 &  & -76 & 0.0753 & 0.9210 & 0.0036 \\
	  &&&&&\\
	& & +76 & 0.0000 & 0.9843 & 0.0157 \\
22.5 (${\rm [Fe/H]}  < -0.6$) & 110 & 0 & 0.1470  & 0.8522 & 0.001 \\
	&  & - 76 & 0.0041 & 0.9920 & 0.0039 \\
\hline
\end{tabular}
\caption{Kinematic group membership probabilities ($P_D$ = thin disc, $P_{TD}$ = thick disc, $P_H$ = halo)  for ULAS~J1316+0755 based on three assumptions of distance: 1) 60~pc corresponding to the target having the mean $M_{J}$ for T6 dwarfs; 2) 38~pc, corresponding to the target being fainter by 1 magnitude as seen for some metal-poor T8 dwarfs; 3) 22~pc corresponding to the target being fainter by 2.25 magnitudes as seen for the T8 subdwarf Wolf 1130C.  Each case has been calculated for plausible radial velocities in the range  $V_{rad} = \pm V_{tan} / \sqrt{2}$~kms${-1}$.
\label{tab:probs}
}
\end{table}

\begin{figure}
\includegraphics[width=8cm,angle=0]{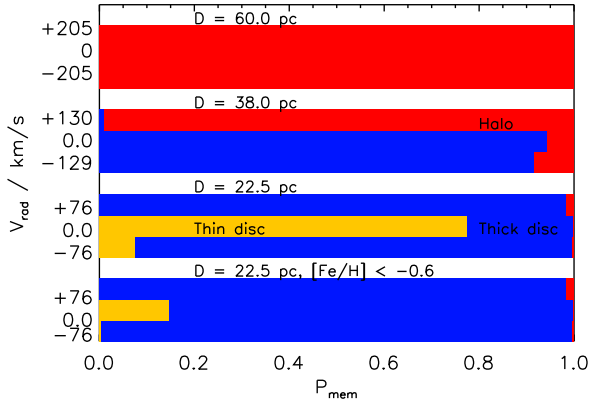}
\caption{A graphical representation of the membership probabilities given in Table~\ref{tab:probs}. The red volumes represent the $P_{H}$, blue $P_{TD}$ and yellow $P_{D}$. The lowest set of probabilities correspond to the 22.5~pc probabilities adjusted for fraction of thin disc stars with $[Fe/H] < -0.6$, see text for explanation.}
\label{fig:memprob}
\end{figure}

Clearly, the assumed distance has a strong impact on the membership we assign to this object. The two larger distance estimates strongly favour thick disc and halo membership, essentially ruling out thin disc membership. The smaller distance favours thin disc membership in the absence of any consideration of metallicity. However, for ULAS~J1316+0755 to be as close as 22.5pc requires it to be as underluminous for its type as Wolf~1130C. We might, reasonably, speculate that this would require ULAS~J1316+0755 to be at least as metal poor as Wolf~1130C, i.e. ${\rm [Fe/H]} < -0.6$. Such metal poor stars are relatively rare in the thin disc, and it is thus appropriate to correct $P_D$ for this fact. \citet{bensby2013b} find that less than 5\% of thin disc stars have ${\rm [Fe/H]} < -0.6$.  On the other hand, this represents a fairly typical metallicity for the thick disc.  We can thus calculate our estimated membership probabilities for this extreme, $D = 22.5$~pc, case by adjusting the normalisation due to the relative populations of thin and thick disc stars for the rarity of thin disc stars with such low-metallicity. When this is taken into account, our adjusted membership probabilities for the smallest distance somewhat favour thick disc membership over the thin disc ($P_{TD} \approx 0.85$).

Although much of this discussion is necessarily speculative at this stage, the balance of probabilities suggests this object is most likely a member of thick disc, and possibly a member of the halo. The $YJHK$ spectrum of this object shows the most extreme deviation from the T6 or T7 standards yet seen, and the deviations are consistent with the expectations for a very low-metallicity object.  We thus argue that ULAS~J1316+0755 should be classified as a T subdwarf, with a spectral type sdT6.5, following the convention established for warmer members of the Galactic thick disc and halo.

\section{Summary}
\label{sec:summ}
We have identified a sdT6 dwarf in the LAS proper motion catalogue presented by \citet{smith2014}. Its high proper motion suggests that it is a member of either the halo or thick disc, and its peculiar spectral morphology is suggestive of very low-metallicity. This object joins a small but growing set of fast moving late-type T dwarfs that are beginning the probe the metallicities below those typical of the Galactic thin disc.  Parallax measurements for this, and the other possible T subdwarfs that are now being discovered, will be essential for distinguishing its membership of the halo or thick disc, and for understanding the impact of low-metallicity on the observed properties of substellar subdwarfs. Since this object is badly blended with a background source in WISE, longer wavelength follow-up observations will also be essential to allow the full SED to be used to distinguish differences in luminosity from flux suppression or enhancement in specific photometric bands. 

\section*{Acknowledgements}
Based on observations made under project A22TAC$\_$96 on the Italian Telescopio Nazionale Galileo (TNG) operated on the island of La Palma by the Fundaci—n Galileo Galilei of the INAF (Istituto Nazionale di Astrofisica) at the Spanish Observatorio del Roque de los Muchachos of the Instituto de Astrofisica de Canarias. Based on observations obtained at the Gemini Observatory,
which is operated by the
Association of Universities for Research in Astronomy, Inc., under a cooperative agreement
with the NSF on behalf of the Gemini partnership: the National Science Foundation (United
States), the
National Research Council (Canada), CONICYT (Chile), the Australian Research Council (Australia),
Minist\'{e}rio da Ci\^{e}ncia e Tecnologia (Brazil)
and Ministerio de Ciencia, Tecnolog\'{i}a e Innovaci\'{o}n Productiva (Argentina).
We would like to acknowledge the support of
the Marie Curie 7th European Community Framework Programme grant n.247593
Interpretation and Parameterization of Extremely Red COOL dwarfs (IPERCOOL)
International Research Staff Exchange Scheme.
This research has made use of the NASA/ IPAC Infrared Science Archive, which is operated by the Jet Propulsion Laboratory, California Institute of Technology, under contract with the National Aeronautics and Space Administration.
This research has made use of the SIMBAD database,
operated at CDS, Strasbourg, France, and has benefited from the SpeX
Prism Spectral Libraries, maintained by Adam Burgasser at
http://www.browndwarfs.org/spexprism.
The authors wish to recognise and acknowledge the very significant cultural role and reverence that the summit of Mauna Kea has always had within the indigenous Hawaiian community.  We are most fortunate to have the opportunity to conduct observations from this mountain.
\bibliographystyle{mn2e}
\bibliography{refs}

\end{document}